\documentclass[sn-mathphys-num]{sn-jnl}

\usepackage{graphicx}%
\usepackage{multirow}%
\usepackage{amsmath,amssymb,amsfonts}%
\usepackage{amsthm}%
\usepackage{mathrsfs}%
\usepackage[title]{appendix}%
\usepackage{xcolor}%
\usepackage{textcomp}%
\usepackage{manyfoot}%
\usepackage{booktabs}%
\usepackage{algorithm}%
\usepackage{algorithmicx}%
\usepackage{algpseudocode}%
\usepackage{listings}%
\usepackage{caption}%

\usepackage[T1]{fontenc}
\usepackage{caption}
\DeclareCaptionLabelFormat{custom}{Fig.~#2 \textbar}
\DeclareCaptionLabelFormat{extended}{Extended Fig.~#2 \textbar}
\theoremstyle{thmstyleone}%
\theoremstyle{thmstyletwo}%

\theoremstyle{thmstylethree}%

\begin{document}

\title[Article Title]{Single-frame super-resolution via Sparse Point Optimization}


\author[1]{\fnm{Xiaofeng} \sur{Zhang}}\email{zhangxf88@mail2.sysu.edu.cn}

\author*[1]{\fnm{Yongsheng} \sur{Huang}}\email{huangysh59@mail.sysu.edu.cn}

\author[2]{\fnm{Jielong} \sur{Yang}}\email{jyang022@e.ntu.edu.sg}

\author[3]{\fnm{Zhili} \sur{Wang}}\email{dywangzl@hfut.edu.cn}

\author[4]{\fnm{Si} \sur{Chen}}\email{chensi000214@gmail.com}

\author*[5]{\fnm{Linbo} \sur{Liu}}\email{liu\_linbo@gzlab.ac.cn}

\author*[1]{\fnm{Xin} \sur{Ge}}\email{ustcgxtc@gmail.com}

\affil[1]{\orgdiv{School of Science}, \orgname{Shenzhen Campus of Sun Yat-sen University}, \orgaddress{\city{Shenzhen}, \postcode{518107}, \country{China}}}

\affil[2]{\orgdiv{College of Information Science and Engineering}, \orgname{Northeastern University}, \orgaddress{\city{Shenyang}, \postcode{110819}, \country{China}}}

\affil[3]{\orgdiv{School of Physics}, \orgname{Hefei University of Technology}, \orgaddress{\city{Anhui}, \postcode{230009}, \country{China}}}

\affil[4]{\orgdiv{Zhongshan Ophthalmic Center}, \orgname{Sun Yat-Sen University}, \orgaddress{\city{Guangdong}, \postcode{510623}, \country{China}}}

\affil[5]{\orgdiv{Guangzhou National Laboratory. No. 9 XingDaoHuanBei Road},  \orgname{Guangzhou International Bio Island}, \orgaddress{\city{Guangzhou}, \postcode{510005}, \country{China}}}


\abstract{Fluorescence microscopy is essential in biological and medical research, providing critical insights into cellular structures. However, limited by optical diffraction and background noise, a substantial amount of hidden information is still unexploited. To address these challenges, we introduce a novel computational method, termed Sparse Point Optimization Theory (SPOT), which accurately localizes fluorescent emitters by solving an optimization problem. Our results demonstrate that SPOT successfully resolves 30 nm fluorescent line pairs, reveals structural details beyond the diffraction limit in both Airyscan and structured illumination microscopy, and outperforms established algorithms in single-molecule localization tasks. This generic method effectively pushes the resolution limit in the presence of noise, and holds great promise for advancing fluorescence microscopy and analysis in cell biology.}




\maketitle

\section{Introduction}\label{sec1}

The diffraction limit fundamentally restricts the resolution of optical microscopes, limiting the ability to resolve nanoscale details of biological structures\cite{Abbe2009}. To address this, super-resolution (SR) microscopy techniques have been developed, enabling researchers to visualize features beyond the diffraction limit\cite{NEICE2010117,Schermelleh2019SuperresolutionMD}.
Among these, methods based on point spread function (PSF) engineering, such as stimulated emission depletion (STED)\cite{hell1994breaking,klar2000fluorescence}, structured illumination microscopy (SIM)\cite{gustafsson2000surpassing,heintzmann2002saturated,gustafsson2005nonlinear}, and image scanning microscopy (ISM)\cite{muller2010image}, have achieved spatial resolutions of approximately 2-4 times better than the diﬀraction limit\cite{kubalova2021comparing}.
Alternatively, single-molecule localization microscopy (SMLM) techniques\cite{balzarotti2017nanometer}, such as STORM\cite{rust2006sub}, PAINT\cite{oi2020live} and PALM\cite{betzig2006imaging}, can overcome the diffraction limit by precisely localizing individual fluorescent molecules, pushing resolution to the order of 10-20 nm\cite{kubalova2021comparing}.

While hardware-based SR techniques have revolutionized optical microscopy,
recent advances in computational imaging suggest that SR can also be
achieved through post-processing, without the need for hardware modifications.
Conventional deconvolution methods such as Wiener or Richardson-Lucy\cite{richardson1972bayesian,lucy1974iterative,lucy1992resolution}, aim
to recover high-frequency information in the Fourier domain, but often suffer from
noise amplification and ill-convergence\cite{tang2015defocused,dey2006richardson,laasmaa2011application,liu2025noise}. In contrast, algorithms such as SRRF
\cite{gustafsson2016fast}, MSSR\cite{torres2022extending}, DPR\cite{zhao2023resolution} and sparse deconvolution\cite{zhao2022sparse} leverage radial fluctuations, mean-shift strategy, pixel reassignment or a priori sparsity-continuity knowledge to computationally extract hidden spatial information. These
methods typically rely on an estimated PSF to model the image formation process
and incorporate prior assumptions for stable reconstruction.

In this work, we introduce sparse point optimization theory (SPOT), a novel computational algorithm that localizes emitting fluorescence points by solving an optimization problem. SPOT leverages sparsity, non-negativity constraints and second-order regularization to realize a SR reconstruction.  Furthermore, to enhance the image quality under low signal-to-noise ratio (SNR) conditions, we propose a modified rolling ball method, termed Segmented Rolling Ball (SRB), which effectively preserves structural details while suppressing noise.
Experimental results demonstrate that SPOT robustly enhances
resolution under various imaging conditions.

\section{Results}\label{sec2}

\subsection{Method execution}\label{subsec1}

SPOT in real space is performed by assuming each pixel as a single-point
emitter. We estimate the brightness of these emitters to derive an intensity
distribution that most closely approximates the ideal object image, while implicitly
imposing the non-negativity constraint on the solution. 
The optimization is carried out using an estimated PSF through a least-squares operation. The transition from high-definition to standard-definition images is handled by a downsampling operation, as described in Equation:
\begin{equation*}
\underset{{x}}{\operatorname{argmin}}\left\{\frac{}{}\|D({p*x})-{f}\|_2^2+(\lambda-1)\|{x}\|_2^2\right\}~~~~~~~~~~s.t.~~x \geq 0,
\label{eq1}
\end{equation*}
where $p$ denotes the upsampled PSF, which is convolved with the guessed high-definition image $x$. The result is then downsampled by operator $D$ to match the matrix size of the acquired standard-definition image $f$. The notation $\left\|\cdot\|_2\right.$ represents the euclidean (second-order) norm.

Our method seeks a solution that minimizes the sum of squared errors, transforming the problem into a least-squares optimization. 
Instead of applying post hoc pixel expansion to the observed image $f$, sub-pixel information is incorporated directly into the optimization by parameterizing the unknown image on a higher-resolution grid (Supplementary Note 6.4).

Although upsampling to a
high-definition image increases the solution dimensionality and improves the
modeling accuracy, it also introduces an underdetermined system, where the
number of unknowns far exceeds the number of equations. This leads to an
ill-conditioned Hessian matrix for the quadratic term\cite{haben2014conditioning}. To alleviate this ill-posed problem, we introduce a second-order regularization term\cite{diffellah2021image}, which serves two main purposes. First, it penalizes isolated high-intensity pixels,
encouraging spatial uniformity in the reconstructed image. Second, it reduces the
condition number of the Hessian matrix, thereby improving numerical stability and
ensuring the uniqueness of the solution. While stronger regularization may lower
the achievable resolution (Supplementary Note 5.2), it remains essential under
realistic imaging conditions, where noise, low contrast, or illumination variation are
inevitable. In ideal noise-free cases, this term should be omitted. However, in
practice, an appropriately selected regularization weight is required for stable and
reliable reconstruction.

In addition, we enforce a non-negativity constraint, which stabilizes the
optimization and improves reconstruction fidelity (Supplementary Note 1.4). This constraint helps suppress
oscillatory artifacts with fluctuating positive and negative values, that commonly
arise during upsampling. Moreover, non-negativity
increases the sparsity of the solution, which is often considered beneficial for
super-resolution reconstruction\cite{zhao2022sparse,xing2020L1}. By restricting the spatial spread of
edge-related artifacts, this constraint also facilitates block-wise image processing
(Supplementary Note 1.5).

Although the SPOT algorithm does not change the SNR of the input image, it
is imperative to suppress noise components, particularly for defocused signals and
background fluorescence (Supplementary Note 2.1). Uniformly distributed noise can
be deconvolved into undesirable artifacts, resulting in cluttered artifacts that obscure real features (Supplementary Note 2.2). To address this, we developed the SRB
algorithm, building upon the conventional rolling ball algorithm\cite{sternberg1983biomedical}. Moreover, to improve multiframe reconstruction under variable PSFs, we
propose the Uniformly Structured Reconstruction (USR) algorithm. Detailed
descriptions of SRB and USR are provided in Supplementary Note 3 \& 4.

\subsection{Resolution Evaluation}\label{subsec4}

To validate the performance of SPOT and its variants (SRB-SPOT and
USR-SRB-SPOT), we conducted tests using a commercial calibration slide
(Argo-SIM, patternE). This slide contains fluorescent line pairs with spacings
incrementally increasing from 0 to 390 nm in 30 nm steps (Supplementary Note 6.1).
Imaging was performed on two microscopes: the Zeiss Elara 7 (in SIM mode) and the
Zeiss LSM 980 (in Airyscan mode)\cite{huff2015airyscan}.

\begin{figure}[!htbp]
\centering
\includegraphics[width=0.9\textwidth]{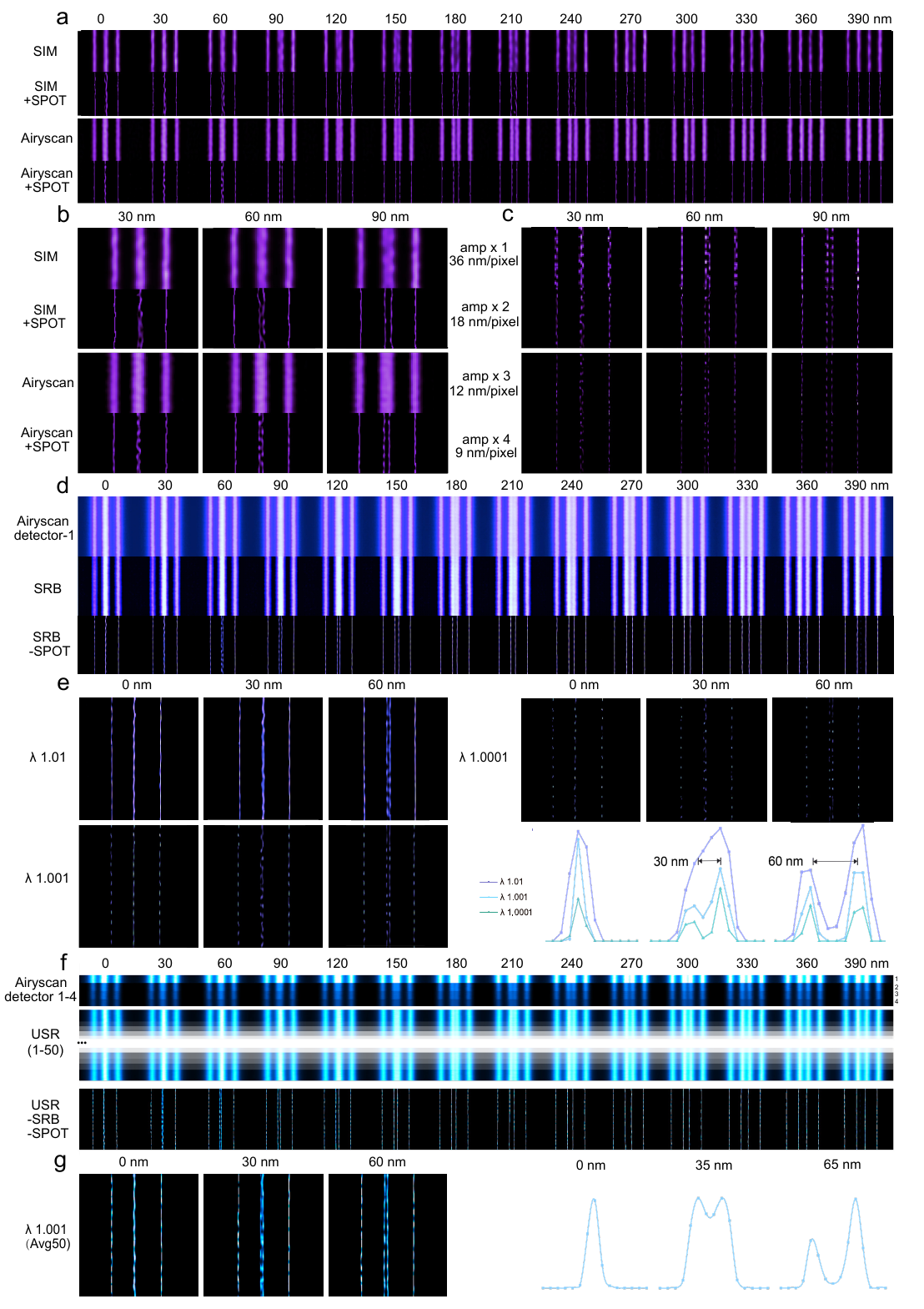}
\caption{
\textbf{Evaluation of SPOT parameter settings and applicability of SRB and USR
using fluorescent line pairs. SPOT achieves a quantitative resolution of 30 nm.}
\textbf{a}, Raw images acquired from SIM and Airyscan microscopy, showing fluorescent line pairs
ranging from 0 to 390 nm, processed with SPOT ($\lambda$ = 1.01).
\textbf{b}, Enlarged views of line pairs at 30, 60, and 90 nm from \textbf{a}.
\textbf{c}, Enlarged views of SPOT-processed Airyscan images
from \textbf{a}, using different pixel amplification factors ($amp$) on the same line pairs ($\lambda$ =
1.001).
\textbf{d}, Raw image acquired from the a single Airyscan center detector (Airyscan detector-1),
processed with SRB, and with SRB followed by SPOT (SRB-SPOT, $\lambda$ = 1.01), showing
line pairs ranging from 0 to 390 nm.
\textbf{e}, Enlarged views of line pairs at 0, 30, and 60 nm.
The $\lambda$ = 1.01 panel is a zoom-in from \textbf{d}, the $\lambda$ = 1.001 and $\lambda$ = 1.0001 panels are reconstructed from the same raw image using SRB-SPOT. The transverse intensity profiles are plotted,
with each point averaged along the vertical direction.
\textbf{f}, Raw images acquired from four Airyscan rings
(Airyscan ring} 1-4), processed with USR to synthesize 50 uniform images. Each was then processed with SRB-SPOT, followed by averaging (USR-SRB-SPOT, $\lambda$ = 1.001).
\textbf{g}, Enlarged views of line pairs at 0, 30, and 60 nm from \textbf{f} (USR-SRB-SPOT, $\lambda$ = 1.001), with corresponding transverse intensity profiles where each point is averaged vertically.
\label{T1}
\end{figure}

In the first test, we applied SPOT to images acquired from both SIM and
Airyscan systems. As shown in Fig.~\ref{T1}a, the original SIM image resolves line pairs at
120 nm spacing, while the Airyscan image resolves down to 150 nm using its built-in
reconstruction (intensity level 6). When SPOT is applied to both types of images
(Fig.~\ref{T1}b), resolution improves to 60 nm.
SPOT also allows controlled pixel-resolution enhancement by adjusting the
amplification factor ($amp$). With $\lambda$ fixed at 1.001, we varied $amp$ to evaluate its
effect on line-pair resolution. As shown in Fig.~\ref{T1}c, $amp$ × 1 corresponds to 36
nm/pixel, $amp$ × 2 to 18 nm/pixel, $amp$ × 3 to 12 nm/pixel, and $amp$ × 4 to 9
nm/pixel. Results show that pixel-resolution expansion significantly improves the
resolution of closely spaced features (e.g., 30 nm), while offering limited benefit for wider spacings (e.g., 90 nm). This suggests that pixel-resolution expansion becomes essential only when the target structures are near or below the pixel size (Supplementary Note 5.3). However, larger $amp$ increases the number of unknowns, making the optimization underdetermined and the reconstructed image sparse.

To evaluate the effectiveness of the SRB algorithm in background noise removal,
we processed a single raw image acquired from the LSM 980 using only the Airyscan
center detector. This raw detector image (first row, Fig.~\ref{T1}d) shows lower resolution and
strong background fluorescence compared to the raw Airyscan image (third row,
Fig.~\ref{T1}a). Direct SPOT processing to this raw detector image is often accompanied by background-induced artifacts (Supplementary Note 2.2). After applying SRB
(second row, Fig.~\ref{T1}d), SPOT successfully resolves structures down to 60 nm (Supplementary Note 6.3), which
is threefold resolution improvement over the detector image (first row, Fig.~\ref{T1}d). As shown in Fig.~\ref{T1}e, 
further reduction of $\lambda$ allows SPOT to resolve structures down to 30
nm (Supplementary Note 6.2). Theoretically, with infinite
bit depth and SNR, the highest reconstructed resolution could be achieved by setting
$\lambda$ = 1. However, as $\lambda$
approaches 1, SPOT-processed images become discrete, as shown in
Fig.~\ref{T1}e. For example, $\lambda$ = 1.001 allows for resolving 30 nm line pairs, but also results in reduced structural continuity. To further validate SPOT's
ability to resolve 30 nm line pairs, we apply it to the publicly available MSSR dataset\cite{torres2022extending}, with results shown in Supplementary Figures 35 and 36.

To process multiple raw images acquired under varying illumination or
detection conditions, we propose the USR method. In Airyscan mode, fluorescence
signals are collected simultaneously by a ring array of multiple detectors. Due to
limited photon counts and varying collection angles, single-detector images often
suffer from low SNR, and vary in brightness and PSFs. In this test, we use only four
single-detector images as inputs (Fig.~\ref{T1}f). USR first transforms these images into a set of 50 synthetic images with uniform PSFs (Supplementary Note 4). Each synthetic image is then denoised by SRB, followed by SPOT processing with $\lambda$ = 1.001. The processed images are averaged to generate a final reconstruction. Fig.~\ref{T1}g shows the reconstructed image and corresponding peak-to-peak measurements for line pairs with spacings of 0, 30, and 60 nm.

Together, these results demonstrate that, under high-SNR conditions, SPOT alone can quantitatively achieve an effective spatial resolution of 30 nm, representing nearly a fourfold improvement over conventional SIM.
Under low-SNR conditions, SRB-SPOT effectively removes background noise and
enhances resolution. When input images vary in illumination or detection conditions, USR-SRB-SPOT still enhances resolution while further improving reconstruction quality.

\subsection{Applications in Subcellular Structural Imaging}\label{subsec5}

\begin{figure}[!htbp]
\centering
\includegraphics[width=0.9\textwidth]{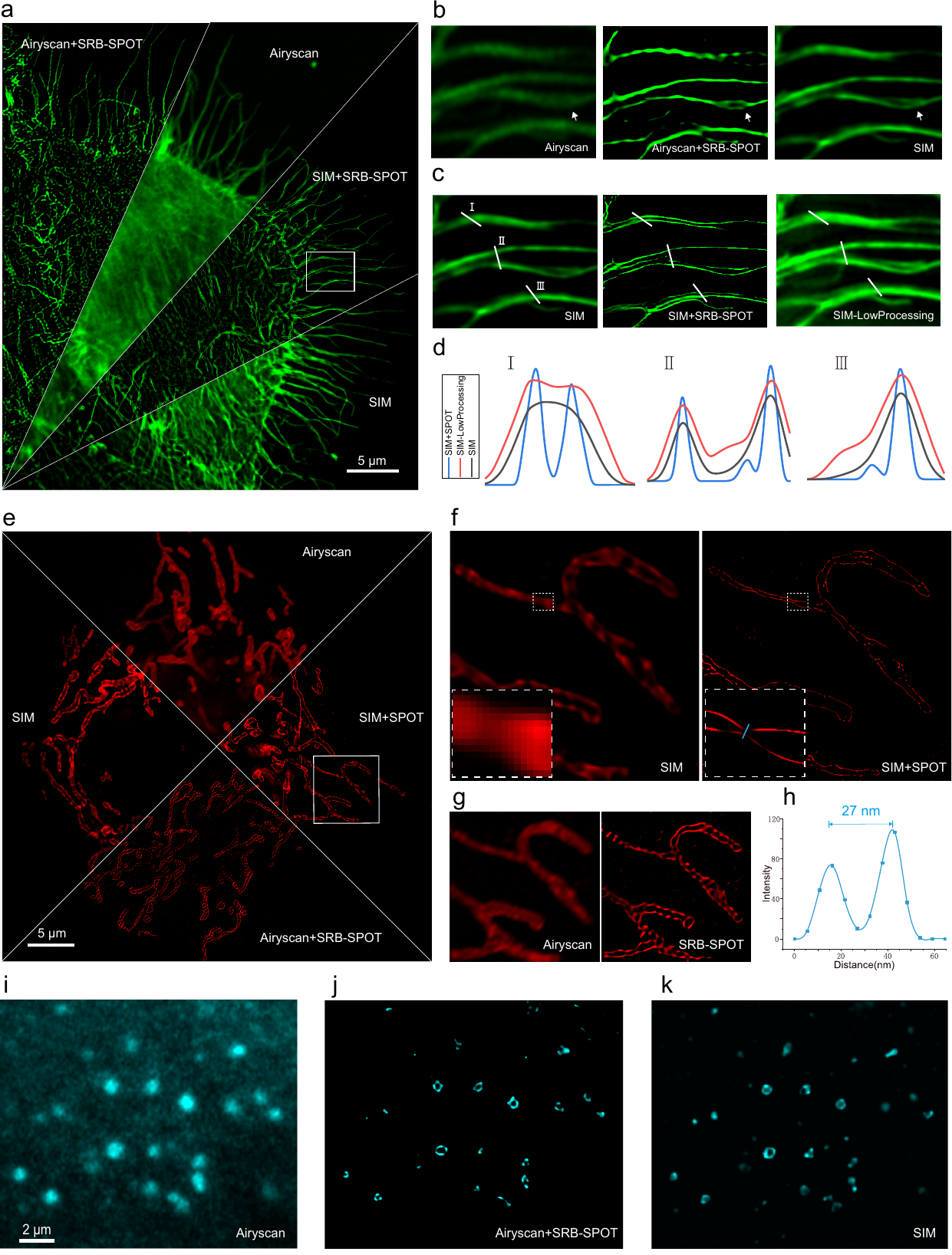}
\caption{
\textbf{Performance of SPOT on subcellular structures. SPOT resolves subcellular structures down to 30 nm.}
\textbf{a}, Raw cytoskeletal images acquired from Airyscan and SIM microscopy, processed with SRB-SPOT.
\textbf{b}, Enlarged views of the white-boxed region in \textbf{a}, comparing Airyscan, SRB-SPOT-processed Airyscan, and SIM (reference). A ring-shaped filopodial structure is indicated by a white arrow.
\textbf{c}, Enlarged views of the same white-boxed region in \textbf{a}, comparing SIM, SRB-SPOT-processed SIM, and SIM low-level processed (reference). 
\textbf{d}, Intensity profiles along the white lines at positions marked by roman numerals (I, II, and III) in \textbf{c}.}
\textbf{e}, Mitochondrial images acquired from Airyscan and SIM microscopy, processed with SRB-SPOT and SPOT, respectively.
\textbf{f}, Enlarged views of the white-boxed region in \textbf{e}, comparing SIM and SPOT-processed SIM.
Enlarged views of the white dashed boxed region in \textbf{f} are shown in the inset.
\textbf{g}, Enlarged views of the same white-boxed region in \textbf{e}, comparing Airyscan and
SRB-SPOT-processed Airyscan.
\textbf{h}, Intensity profile along the blue line in \textbf{f}, showing a peak-to-peak distance of 27 nm.
\textbf{i-k}, Lysosomal images acquired from Airyscan (\textbf{i}), Airyscan processed with SRB-SPOT (\textbf{j}), and SIM for comparison (\textbf{k}).\label{T2}
\end{figure}

To evaluate the broad applicability of the SPOT, we applied it to raw images of
cytoskeleton, mitochondria, and lysosome. Detailed microscope settings and sample
preparation are described in the \textquotedbl Methods\textquotedbl~section. Fig.~\ref{T2}a shows an overview of the
cytoskeletal structures, with magnified views of the white-box region shown in
Fig.~\ref{T2}b and \ref{T2}c. The Airyscan raw image (left, Fig.~\ref{T2}b) shows lower resolution and
SNR than the SIM raw image (right, Fig.~\ref{T2}b). After SRB-SPOT processing
(middle, Fig.~\ref{T2}b), a ring-like filopodial structure is revealed (white arrow), consistent
with the feature observed in the SIM raw image (right, Fig.~\ref{T2}b), confirming its
actual presence. SIM raw image (left, Fig.~\ref{T2}c) is further processed with SRB-SPOT,
revealing additional fine structural details indicated by roman numerals (middle,
Fig.~\ref{T2}c). To verify their existence, we used a low-level processing version of the SIM
image (right, Fig.~\ref{T2}c), which retains defocus information from adjacent planes.
By comparing the middle and right images in Fig.~\ref{T2}c, we confirm that these resolved structures (middle, Fig.~\ref{T2}c)
are not artifacts introduced by SPOT processing. 
The intensity profile of SIM-LowProcessing also suggests the presence of these separated structures (Fig. 2d).
This cross-validation demonstrates
that SPOT can effectively resolve real features, rather than simply enhancing image
sharpness.

Fig.~\ref{T2}e shows an overview of mitochondrial structures. The SRB-SPOT processed
Airyscan image (bottom, Fig.~\ref{T2}e) shows that defocused signals are effectively
removed, in contrast to the Airyscan raw image (top, Fig.~\ref{T2}e). Magnified views of the white-box region in Fig.~\ref{T2}e are shown in Fig.~\ref{T2}f
and \ref{T2}g. In Fig.~\ref{T2}f, SIM images before and after SPOT processing are compared. The dashed-box region is further enlarged in the inset, and the corresponding intensity profile along the blue line is plotted in Fig.~\ref{T2}h, revealing a peak-to-peak distance of 27 nm. Fig.~\ref{T2}g shows the corresponding comparison for Airyscan images before and after SRB-SPOT processing.

Fig.~\ref{T2}i-\ref{T2}k show the results for lysosomes. After processing, the Airyscan raw image exhibits a significant improvement in resolution, as shown in Fig.~\ref{T2}j. The SIM image in Fig.~\ref{T2}k serves as the ground truth for the SRB-SPOT–processed Airyscan result in Fig.~\ref{T2}j. Applying SPOT to the SIM image can further enhance spatial resolution. The resulting structures, shown in Supplementary Fig. 34, clearly reveal sub-pixel features beyond the original resolution limit.

\subsection{Resolution Comparison on Public Datasets}\label{subsec6}

\begin{figure}[!htbp]
\centering
\includegraphics[width=0.9\textwidth]{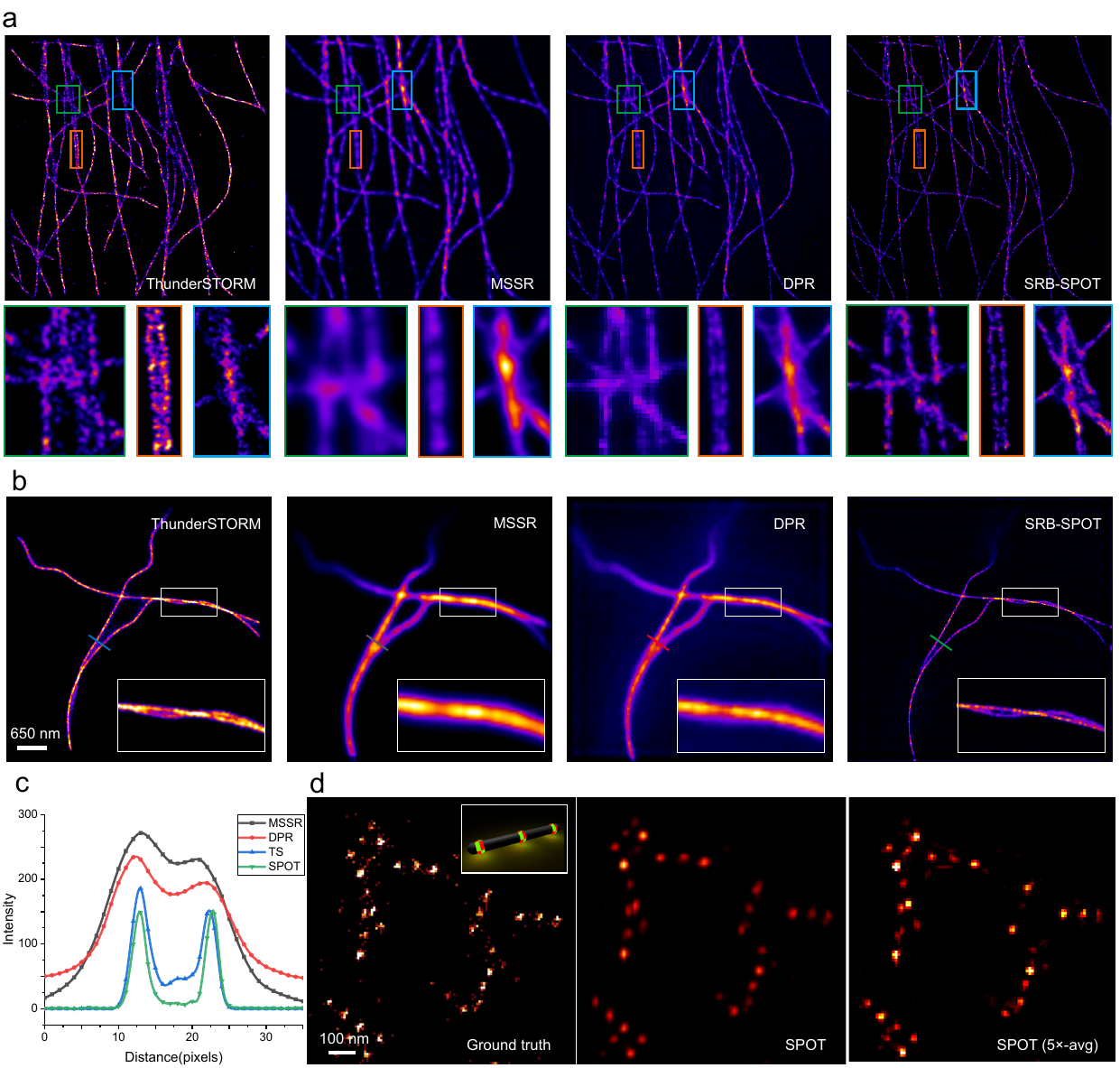}
\caption{
\textbf{Resolution enhancement of single-molecule localization microscopy images using SPOT.}
\textbf{a}, Processing results of 500 densely populated SMLM images using ThunderSTORM, MSSR, DPR, and SRB-SPOT (left to right). Colored boxes indicate enlarged views of corresponding regions.
\textbf{b}, Processing results
of 2500 SMLM images using the same four algorithms, with enlarged views shown in insets.
\textbf{c}, Pixel intensity profiles along the colored lines in \textbf{b}.
\textbf{d}, SMLM results of 80 nm fluorescent nanorods from GATTAquant. Left: ground truth image from the
official website, with a rendered nanorod in the inset. Middle: SPOT applied to each frame. Right: every five frames averaged first, followed by SPOT.
}\label{T3}
\end{figure}

Although SPOT is primarily designed for single-frame processing, it can be
extended to multi-frame datasets such as SMLM through frame averaging. Here, we
use ThunderSTORM, a widely used SMLM reconstruction tool, as a benchmark. In Fig.~\ref{T3}a and \ref{T3}b, we compare SPOT with MSSR\cite{torres2022extending} and DPR\cite{zhao2023resolution} on SMLM datasets,
applying each method to raw frame sequences followed by averaging. Fig.~\ref{T3}a shows
results from a public dataset of 500 diffraction-limited images of tubulin-labeled
microtubules at high fluorophore density\cite{sage2015quantitative}. ThunderSTORM performs well in
regions with low fluorophore density (orange box, first column, Fig.~\ref{T3}a) but
struggles in regions where microtubules intersect (green and blue boxes, first
column, Fig.~\ref{T3}a). MSSR reconstructs the overall filamentous structures but fails to
recover fine details (second column, Fig.~\ref{T3}a). DPR resolves structures in the
orange-box region but loses performance in the densely entangled regions (third
column, Fig.~\ref{T3}a). In contrast, SPOT successfully resolves microtubule networks across
all colored-box regions (fourth column, Fig.~\ref{T3}a).

Fig.~\ref{T3}b shows results from another public SMLM dataset, which contains 2500
frames with low fluorophore density. MSSR produces smooth reconstructions with
low background noise but at the cost of lower resolution. DPR achieves higher resolution than MSSR but introduces increased background noise. In contrast, SPOT
clearly distinguishes two closely intertwined microtubules, a feature also verified
by ThunderSTORM but unresolved by MSSR and DPR (white box, Fig.~\ref{T3}b). Pixel
intensity profiles along the colored lines in Fig.~\ref{T3}b, shown in Fig.~\ref{T3}c, demonstrating
SPOT’s ability in resolving fine structures with reduced background fluorescence. 

To further evaluate the performance of the SPOT in SMLM, we processed a
publicly available dataset from the GATTAquant website\footnote{\url{https://www.gattaquant.com/downloads}}. This dataset consists of
fluorescent nanorods with emission points spaced 80 nm apart. As shown in Fig.\ref{T3}d, SPOT effectively resolves individual nanorods with reduced background
noise, outperforming the reference image provided by the website. To simulate higher temporal resolution in SMLM, we averaged every five images into a single frame to increase the emitter density.
Even under this condition, SPOT maintains comparable spatial resolution
(Supplementary Note 2.3). This suggests that our methods can potentially achieve a
fivefold improvement in temporal resolution, marking a promising advance for SMLM
applications\cite{lelek2021single,speiser2021deep}.

\subsection{Quantitative Resolution Evaluation via rFRC Analysis}\label{subsec7}

\begin{figure}[!htbp]
\centering
\includegraphics[width=0.9\textwidth]{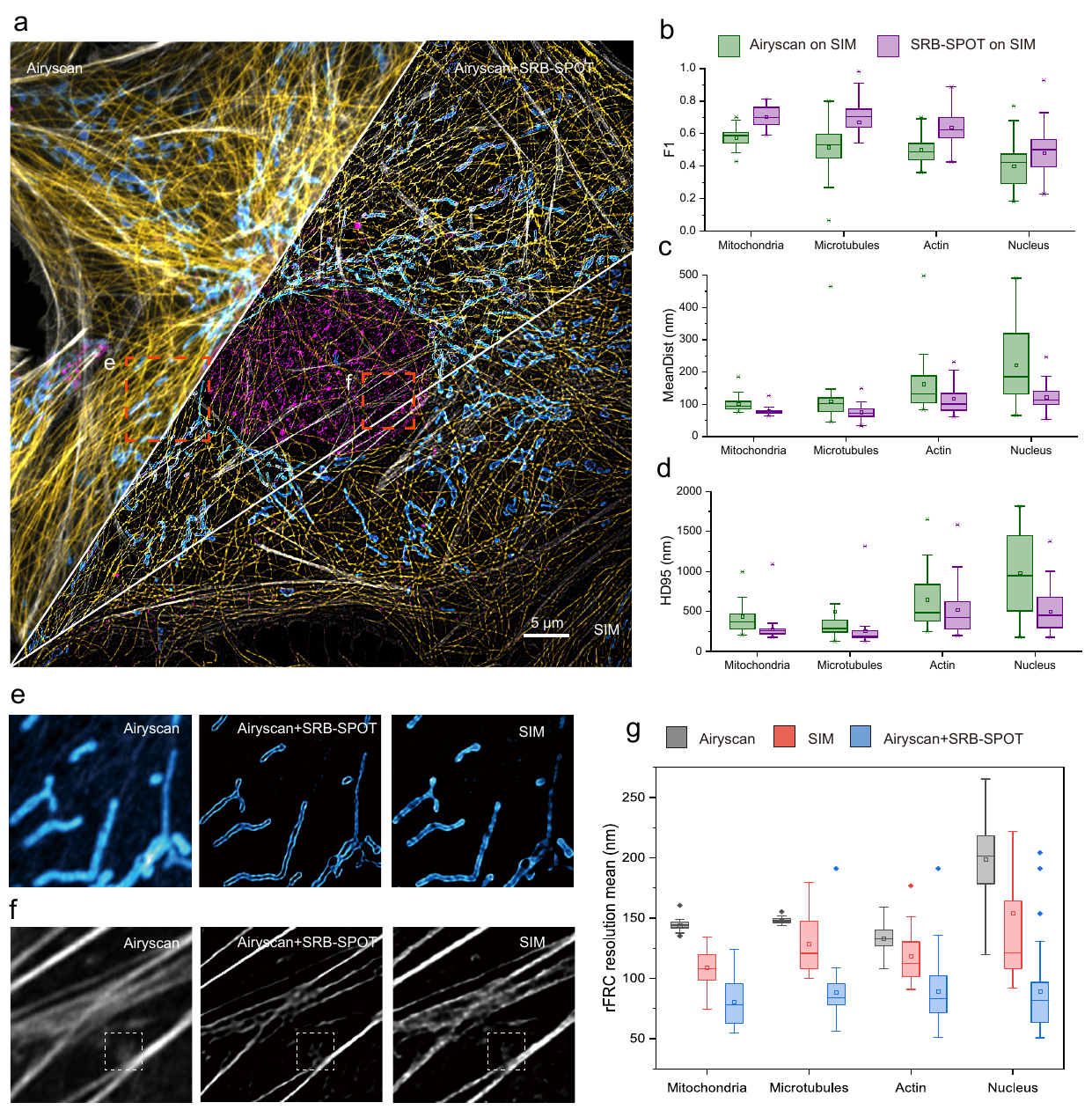}
\caption{
\textbf{Multichannel super-resolution and quantitative analysis of SPOT.}
\textbf{a}, Four-color image of a COS-7 cell showing
mitochondria (cyan), microtubules (orange), actin (gray), and nucleus
(magenta). Images were acquired from Airyscan and SIM microscopy, processed with SRB-SPOT.
\textbf{b–d}, Quantitative evaluation of structural fidelity against the SIM reference. \textbf{b}, Skeleton-based F1 score. \textbf{c}, Mean skeleton distance (MeanDist). \textbf{d}, 95th percentile Hausdorff distance (HD95). Metrics were computed on skeletonized structures using an identical processing pipeline with tolerance-based matching. Higher F1 and lower MeanDist and HD95 indicate improved agreement with the SIM reference.
\textbf{e and f}, Enlarged views of the regions indicated dashed orange boxes in \textbf{a}, showing  mitochondria (\textbf{e}) and actin filaments (\textbf{f}), respectively. Each panel shows Airyscan, SRB-SPOT processed Airyscan, and SIM for comparison.
}
\textbf{g}, Average resolution of different subcellular structures
estimated by rFRC. Each box plot includes 49 images. Gray,
red and blue boxes represent Airyscan, SIM and SRB-SPOT-processed Airyscan, respectively. Box plot elements: horizontal line, median; squares, mean; box limits, interquartile range (IQR) 25\%-75\%; whiskers, 1.5 IQR; diamonds, outliers.\label{T4}
\end{figure}

To quantitatively evaluate the performance of SPOT on biological samples, we applied it to multichannel fluorescence images of an Origin Simian-7 (COS-7) cell, acquired from Airyscan and SIM. Raw Airyscan images exhibit substantial background fluorescence, partly due to out-of-focus contributions. After SRB-SPOT processing, background fluorescence is eﬀectively suppressed and structural details become clearer (Fig.~\ref{T4}a).

Structural fidelity was quantified against the SIM reference using skeleton-based metrics, including F1 score, mean skeleton distance (MeanDist), and 95th percentile Hausdorff distance (HD95) (Fig.~\ref{T4}b–d). SRB-SPOT consistently achieves higher F1 scores and lower MeanDist and HD95 across organelles, indicating improved structural recovery with reduced geometric deviation.

Representative enlarged views of mitochondria and actin filaments (Fig.~\ref{T4}e,f) further demonstrate finer structural details after SRB-SPOT processing, together with effective background suppression. Notably, even extremely weak signals (Fig.~\ref{T4}f, dashed box) are faithfully preserved after reconstruction, without structural loss.

Finally, resolution gains were further evaluated by rolling Fourier ring correlation (rFRC) analysis\cite{nieuwenhuizen2013measuring,zhao2023quantitatively}. SRB-SPOT achieves an approximately twofold improvement across subcellular structures. These results demonstrate that SRB-SPOT enhances contrast, suppresses background, and improves eﬀective resolution while preserving structural fidelity.

\section{Discussion}\label{4}

Conventional deconvolution methods such as Wiener or Richardson-Lucy (RL),
often amplify noise during the reconstruction (Supplementary Note 1.2). In
contrast, SPOT formulates image reconstruction as a least-squares optimization,
where a non-negativity constraint effectively suppresses artifact propagation
(Supplementary Note 1.4). To address the high computational cost when processing
large-scale images, SPOT uses a block-wise processing strategy (Supplementary
Note 1.5). When the block size exceeds four times the full width at half maximum
(FWHM) of the PSF, the reconstructed results closely match those from full-image
processing, while computation speed is significantly accelerated (Supplementary Note 5.3).

Like many SR methods, SPOT requires prior knowledge of the PSF’s FWHM.
Although theoretical estimates can be derived from system parameters\cite{zhang2007gaussian}, built-in
post-processing operations in SR microscopy sometimes make the actual PSF
unpredictable. In practice, the FWHM of fine structures in raw images can also be
directly measured as an initial estimate. For SPOT to achieve a resolution of
120 nm, the PSF estimate can tolerate up to 30\% deviation. 
However, for a finer resolution of 30 nm, the tolerance drops to 5\% (Supplementary Note 5.1).
Additionally, SPOT requires tuning a regularization parameter $\lambda$, which balances structural continuity and achievable resolution (Supplementary Note 5.2). A
detailed parameter tuning workflow is provided in Supplementary Note 5.4. Sub-pixel resolution arises from solving the reconstruction in a higher-dimensional space as an underdetermined problem, rather than from simple image-domain interpolation. Under super-resolution conditions, this formulation achieves superior quantitative performance compared to conventional image upsampling followed by SPOT with amp = 1 (Supplementary Note 6.4).

In this work, we present a generalizable framework that realizes SR reconstruction across different imaging modalities. Unlike task-specific SR methods, our method
provides a unified solution for both high- and low-SNR images (Supplementary
Note 2.5). For high-SNR images, SPOT achieves 30 nm resolution, the highest
reported for Airyscan and SIM to our knowledge. 
In low-SNR cases, we introduce a
preprocessing module (SRB) to suppress background noise, followed by SPOT. For
multiple captured images, we propose the USR algorithm to normalize images and
generate an image stack, fully leveraging all image information to achieve optimal
resolution. For SMLM datasets, our method achieves competitive resolution even in
dense fluorescence labeling. This modular strategy enables SPOT to be flexibly
adapted to various imaging modalities, including Airyscan, SIM, SMLM and TIRF
(Supplementary Note 7), consistently improving resolution and structure clarity.
Overall, our framework offers a practical and scalable approach for fluorescence
image enhancement.

\bibliography{sn-bibliography}

\clearpage

 \section*{Methods}

\noindent\textbf{SPOT Algorithm.}  
A flowchart our proposed algorithm is shown in Supplementary Note 1.3,
begins with optional preprocessing modules: USR and SRB. Detailed workflows for SRB and USR
are described in Supplementary Notes 3 and 4, respectively. USR is selectively applied to
multi-frame datasets with heterogeneous detection responses, such as those from multi-detector
systems (e.g., Airyscan). For imaging systems with consistent and uniform responses, such as
single-frame images or SMLM data, this step can be skipped. SRB is applied only when background noise denoising is required.
SPOT is applied to preprocessed images or raw images. For large images, apply block-wise processing. To mitigate boundary artifacts, ensure each sub-image ($Img_b$) is at least four times the PSF’s FWHM (Supplementary Note 1.5). The processed sub-images are then recombined into a full-size image. For each sub-image, the optimization problem is formulated as follows. A high-definition matrix $x$ of
size $(Img_b\times amp) \times (Img_b\times amp)$ is initialized as an all-ones matrix. This matrix is convolved with a PSF kernel from a theoretically
estimated PSF to produce a convolved image $H_1$.
$H_1$ is then corrected by edge compensation to produce $H_2$ (Supplementary Note 1.5). Next, $H_2$ is
flattened and squared, and its main diagonal is scaled by a regularization factor $\lambda$ to form the
Hessian matrix $H_3$. Concurrently, $H_2$ is downsampled by block averaging over an $amp\times amp$ grid to
form D($H_2$). The coefficient of the first-order term is computed as the product of D($H_2$) and the
negative $Img_b$. The final inputs to the solver are the triplet {$Img_b$, $H_3$, D($H_2$)}. The optimization is solved under non-negativity constraints using either MATLAB’s $quadprog$ function or the Gurobi
solver. For multi-frame datasets, each frame is processed independently. Outputs include individual
reconstructions, an averaged image, and a variance image. Detailed parameter settings are
provided in Supplementary Note 5.4.

\bigskip
\noindent\textbf{Microscope settings.}  
The Airyscan image in Fig.~\ref{T1} was acquired from a Zeiss LSM 980 confocal microscope equipped with a Plan-Apochromat 63$\times$/1.40 Oil DIC M27 objective. Excitation was performed with a 405 nm laser at 16-bit depth. Emission was collected from 300–735 nm using a GaAsP-PMT detector, with a gain of 650 V and digital gain of 1. Imaging was conducted in Airyscan mode with a pinhole size of 5.00 AU, a scan zoom of 3.3, a dwell time of 7.26 µs, and a sampling factor of 2.00. Each frame took 2 minutes 40 seconds, with 8$\times$ line averaging and mean-based noise reduction.

The SIM image in Fig.~\ref{T1} was acquired from a Zeiss Elyra7 microscope, also equipped with a Plan-Apochromat 63$\times$/1.40 Oil DIC M27 objective. A 405 nm laser was used for
excitation, and emission was collected from 420–480 nm at 16-bit depth.

The Airyscan image in Fig.~\ref{T2} was acquired from a Zeiss LSM 800 microscope. Cytoskeleton and mitochondria were captured at 16-bit depth, while lysosomes were captured at 8-bit. Excitation wavelengths were 561 nm for cytoskeleton and lysosome and 488 nm for mitochondria. Emission was collected over 450–700 nm for cytoskeleton and lysosome 489–531 nm for mitochondria. Detector gains were set to 850 V for cytoskeleton and lysosome, and 800 V for mitochondria. Scans were unidirectional for cytoskeleton and mitochondria, and bidirectional for lysosome with single applied.

The SIM image in Fig.~\ref{T2} was acquired from a HIS-SIM microscope equipped with a 100$\times$/1.50 Oil immersion objective. Data were collected at 16-bit depth. The raw image represents the default output of the system, which includes a built-in Sparse Deconvolution step set to a minimal intensity level of 1\cite{zhao2022sparse}.

The Airyscan image in Fig.~\ref{T4} acquired from a Zeiss LSM 980 microscope.
Excitation was performed with laser lines at 405 nm (nucleus), 488 nm
(microfilaments), 543 nm (microtubules), and 639 nm (mitochondria). Emissions were
collected in designated spectral windows using a GaAsP-PMT detector. Each
fluorophore was imaged with its specific pinhole size and dwell time. Scanning was
performed in unidirectional mode with a zoom factor of 1.7 and a sampling factor of
2.00.

The SIM image in Fig.~\ref{T4} was from a Zeiss Elyra 7 microscope. Excitation wavelengths were 405 nm, 488 nm, 561 nm, and 642 nm. Emission windows were customized for each fluorophore. Image acquisition used 4$\times$ line averaging.

\bigskip
\noindent\textbf{Cell sample preparation.}  
For Fig.~\ref{T2}, C2C12 cells were cultured on coverslips, fixed with 3\% paraformaldehyde for 30 minutes, and permeabilized with 0.05\% Triton X-100 for 10 minutes. Mitochondria were labeled with an anti-TOM20 primary antibody (Abclonal, A19403, 1:200) at 4°C overnight, followed by a DyLight™ 488-conjugated Goat Anti-Rabbit secondary antibody (Invitrogen, 1:200). For lysosomes, anti-LAMP1 (DSHB, 1D4B, 1:200) and DyLight™ 546-conjugated goat anti-rat secondary antibody (Invitrogen, 1:200) were used. Actin filaments were stained with Phalloidin Peptide (Sigma, P1951, 1:1000). Samples were mounted after staining.

For Fig.~\ref{T4}, U2OS, HeLa, and COS-7 cells were fixed with 4\% paraformaldehyde, permeabilized with 0.1\% Triton X-100, and blocked with 10\% BSA. Primary antibodies included TOM20 (Proteintech, 11802-1-AP, 1:100) and $\alpha$-tubulin (Invitrogen, A48264, 1:200). Secondary antibodies included CoraLite™ 647 Goat Anti-Rabbit (Proteintech, SA00014-9, 1:500) and Alexa Fluor™ Plus 555 Goat Anti-Rat (Invitrogen, A48263, 1:1000). Hoechst dye and Phalloidin (Sigma, P1951) were used to label the nucleus and actin filaments, respectively.

\bigskip
\noindent\textbf{SPOT Algorithm Implementation and PC Configuration.}
The SPOT algorithm was compiled and executed in MATLAB on Windows-based systems. Two PC configurations were used for testing:

\bigskip
\noindent\textbf{Test Configuration 1:}
\begin{itemize}
    \item \textbf{Operating system:} Windows 11 Pro
    \item \textbf{Processor:} 12th Gen Intel Core i5-12400F @ 2.50 GHz
    \item \textbf{Memory:} 16.0 GB (15.9 GB available)
\end{itemize}

\noindent\textbf{Test Configuration 2:}
\begin{itemize}
    \item \textbf{Operating system:} Windows Server 2022 
    \item \textbf{Processor:} AMD EPYC 7773X 64-Core Processor 2.20 @ GHz (2 processors)  
    \item \textbf{Memory:} 512 GB (512 GB available)
\end{itemize}

In Configuration 1, the second-order optimization problem was solved using MATLAB's built-in \texttt{quadprog} function. The solver was configured to use the Interior-Point Method to address the convex quadratic programming problem.

In Configuration 2, the second-order optimization function calculations are performed using the external Gurobi\cite{gurobi} solver, optimized for efficient utilization of multi-core processors. Notably,
the choice of solver (e.g., MATLAB’s \texttt{quadprog} or Gurobi) does not affect the final optimization results.

The SRB algorithm was developed by modifying the rolling ball algorithm provided by Fiji, as detailed in the \textit{MIJ: Running ImageJ and Fiji within Matlab}.\footnote{\url{https://www.mathworks.cn/matlabcentral/fileexchange/47545-mij-running-imagej-and-fiji-within-matlab}}

\bigskip
\noindent\textbf{Other algorithms.} ThunderSTORM was executed in Fiji program\cite{ovesny2014thunderstorm}. The \textquotedbl Multi-emitter fitting analysis\textquotedbl~option was enabled, and the \textquotedbl Maximum of molecules per fitting region\textquotedbl~was set to 5.

DPR (Deblurring by Pixel Reassignment)\cite{zhao2023resolution} enhances image resolution by reassigning pixel intensities based on local gradients, allowing the distinction of closely spaced points even beyond the diffraction limit. This method operates in real space, avoiding noise amplification
common in conventional deconvolution methods. It also ensures intensity preservation without introducing negative values or requiring a highly accurate PSF. DPR was run in MATLAB, with an intensity level of DPR2. The PSF was determined through manual optimization.

The MSSR (Mean-Shift SR)\cite{torres2022extending} is based on the Mean-Shift theory and enhances resolution by refining the spatial distribution of fluorescence signals. It
computes the magnitude of local Mean-Shift vectors to sharpen and redistribute
signal intensities, achieving resolution beyond the diffraction limit. MSSR performs
well at both low and high fluorophore densities and is applicable to single images or
temporal stacks. MSSR was run in the Fiji with \textquotedbl Order\textquotedbl~parameter is set to 1, and the \textquotedbl Interpolation Type\textquotedbl~is set to Bicubic.

Sparse Deconvolution\cite{zhao2022sparse} in SR fluorescence microscopy leverages prior knowledge about the sparsity and continuity of biological structures to enhance image resolution. Techniques such as Sparse Structured Illumination Microscopy (Sparse-SIM) can nearly double the resolution while maintaining high temporal resolution. This methodextracts high-frequency details while suppressing background
and artifacts, even under low-SNR conditions. In Extended Fig.~\ref{T5}, Sparse
Deconvolution was performed using MATLAB code version 1.0.3. The images in Extended Fig.~\ref{T6} are released by theauthors of the Sparse Deconvolution method.

rFRC\cite{nieuwenhuizen2013measuring,zhao2023quantitatively} (rolling Fourier ring correlation) algorithm evaluates image resolution by calculating local correlations between two independent reconstructions in the Fourier domain. By rolling a window across the image, it generates pixel-wise resolution mapping, enabling uncertainty detection at the SR scale.rFRC was run in Fiji using the 3-sigma curve criterion with a block size of 64. For Fig.~\ref{T4} was divided into into 49 sub-images. Statistical plotting was carried out using Origin 2021.

Decorrelation analysis\cite{descloux2019parameter} provides a parameter-free method for resolution
estimation on a single image. It operates in the Fourier domain, applying
normalization and cross-correlation with binary frequency masks to determine the
highest resolvable spatial frequency. Resolution is determined by the
local maximum of the decorrelation function, eliminating the need for user-defined
parameters. This approach enables robust, automated approach supports real-time
resolution evaluation across various imaging modalities, including SR microscopy. In Extended Fig.~\ref{T5}d, decorrelation analysis was used for resolution estimation.

\bigskip
\noindent\textbf{Structural fidelity metrics.}
To quantitatively evaluate the structural agreement between reconstructed images and the higher-resolution SIM reference, three skeleton-based metrics were employed: the F1 score, the mean skeleton distance (MeanDist), and the 95th percentile Hausdorff distance (HD95). These metrics assess complementary aspects of structural fidelity, including the accuracy of feature recovery and the geometric consistency of filamentous structures.

Prior to metric computation, all images were locally aligned using a rigid registration procedure to minimize residual translational offsets. Each image was then binarized using an identical thresholding pipeline based on Otsu’s method, followed by removal of small connected components to suppress noise. Skeleton representations were subsequently extracted using iterative morphological thinning (\texttt{bwmorph} with the \textquotedbl skel\textquotedbl~option in MATLAB), ensuring that comparisons were based on the underlying structural topology rather than intensity variations.

Skeleton-based F1 score.
The F1 score quantifies the overlap between the reconstructed skeleton and the SIM-derived reference while allowing for small spatial tolerances to account for residual registration errors and biological variability. Let $S_{\mathrm{rec}}$ and $S_{\mathrm{SIM}}$ denote the sets of skeleton pixels from the reconstructed image and the SIM reference, respectively. A skeleton pixel in $S_{\mathrm{rec}}$ is considered a true positive (TP) if its Euclidean distance to the nearest pixel in $S_{\mathrm{SIM}}$ is less than or equal to a predefined tolerance radius $r$. Pixels in $S_{\mathrm{rec}}$ without a corresponding match are counted as false positives (FP), while unmatched pixels in $S_{\mathrm{SIM}}$ are treated as false negatives (FN). Precision and recall are defined as:
\begin{equation}
\mathrm{Precision} = \frac{\mathrm{TP}}{\mathrm{TP} + \mathrm{FP}}, \qquad
\mathrm{Recall} = \frac{\mathrm{TP}}{\mathrm{TP} + \mathrm{FN}},
\end{equation}
and the F1 score is computed as:
\begin{equation}
F_1 = \frac{2 \times \mathrm{Precision} \times \mathrm{Recall}}
{\mathrm{Precision} + \mathrm{Recall}}.
\end{equation}
Higher F1 values indicate improved recovery of structural features with fewer false detections and omissions.

Mean skeleton distance (MeanDist). 
The MeanDist metric evaluates the average geometric deviation between two skeletons. Using Euclidean distance transforms, the nearest-neighbor distances from each pixel in $S_{\mathrm{rec}}$ to $S_{\mathrm{SIM}}$, and vice versa, are computed. The symmetric mean skeleton distance is then defined as
\begin{equation}
\mathrm{MeanDist} = \frac{1}{2}
\left(
\frac{1}{|S_{\mathrm{rec}}|}
\sum_{p \in S_{\mathrm{rec}}} d(p, S_{\mathrm{SIM}})
+
\frac{1}{|S_{\mathrm{SIM}}|}
\sum_{q \in S_{\mathrm{SIM}}} d(q, S_{\mathrm{rec}})
\right),
\end{equation}
where $d(p, S)$ denotes the Euclidean distance from pixel $p$ to the closest pixel in set $S$. Lower MeanDist values indicate better spatial agreement with the reference structure.

95th percentile Hausdorff distance (HD95).
To assess the worst-case structural deviation while remaining robust to outliers, the 95th percentile Hausdorff distance was used instead of the conventional maximum Hausdorff distance. Let $D_{rec\rightarrow SIM}$ and $D_{SIM\rightarrow rec}$ denote the sets of nearest-neighbor distances from $S_{\mathrm{rec}}$ to $S_{\mathrm{SIM}}$ and vice versa. The HD95 metric is defined as
\begin{equation}
\mathrm{HD95} = \max \left(
\mathrm{percentile}_{95}(D_{rec\rightarrow SIM}),
\mathrm{percentile}_{95}(D_{SIM\rightarrow rec})
\right).
\end{equation}
Lower HD95 values indicate improved structural consistency with the SIM reference while reducing sensitivity to isolated mismatches.

For all experiments, these metrics were computed on matched image patches to ensure fair comparisons between imaging modalities. Distances measured in pixels were converted to physical units (nanometers) using the corresponding pixel size of each dataset. Statistical analyses and box plots were generated using Origin 2021.

\bigskip
\noindent\textbf{PSF Generation in SPOT.}
SPOT provides two modes for PSF generation: a Gaussian function fit and a Bessel function fit.The selection between these two PSF models can be adjusted in the code, with the default set to the Bessel function fit. 

The Gaussian PSF model is defined as\cite{zhang2007gaussian}:

\begin{equation}
\mathrm{{PSF}_G}=\exp \left(-\frac{x^2+y^2}{2 \sigma^2}\right)
\end{equation}

where $\sigma$ denotes the standard deviation of the Gaussian PSF, which determines its effective spatial extent. The function values are computed within a grid whose side length is ${PSF} \times {amp}$ , and these values are integrated over each grid cell to obtain the final PSF generation matrix.

The expression for the Bessel function fit PSF is\cite{ahi2017mathematical}:

\begin{equation}
\mathrm{{PSF}_{B}}=\left[2 \frac{J_1\left(\rho\right)}{\rho}\right]^2
\end{equation}

 where $J1(\cdot)$ is the first-order Bessel function and $\rho$ is the normalized radial
coordinate. The function values are computed on a grid whose spatial extent reaches the second minimum of the Airy pattern, and the values are integrated over each grid cell to obtain the final PSF generation matrix.Compared to the Gaussian function fit PSF, the Bessel function fit PSF retains the secondary peak, providing a better approximation for certain optical systems.

\clearpage

\section*{Acknowledgements}\label{ACK}
This work was supported in part by the Guangzhou National Laboratory (GZNL2025C03014), Guangdong Basic and Applied Basic Research foundation (2023A1515011289), Guangdong Provincial Key Laboratory of Advanced Particle Detection Technology (2024B1212010005), Guangdong Provincial Key Laboratory of Gamma-Gamma Collider and Its Comprehensive Applications (2024KSYS001). 
We sincerely thank the Shenzhen Brain Science Infrastructure for their technical support and assistance in this study.

\setcounter{figure}{0}
\captionsetup[figure]{labelformat=extended, labelsep=space, labelfont=bf,
    format=plain, textfont=normalfont}

\begin{figure}[!htbp]
\centering
\includegraphics[width=0.9\textwidth]{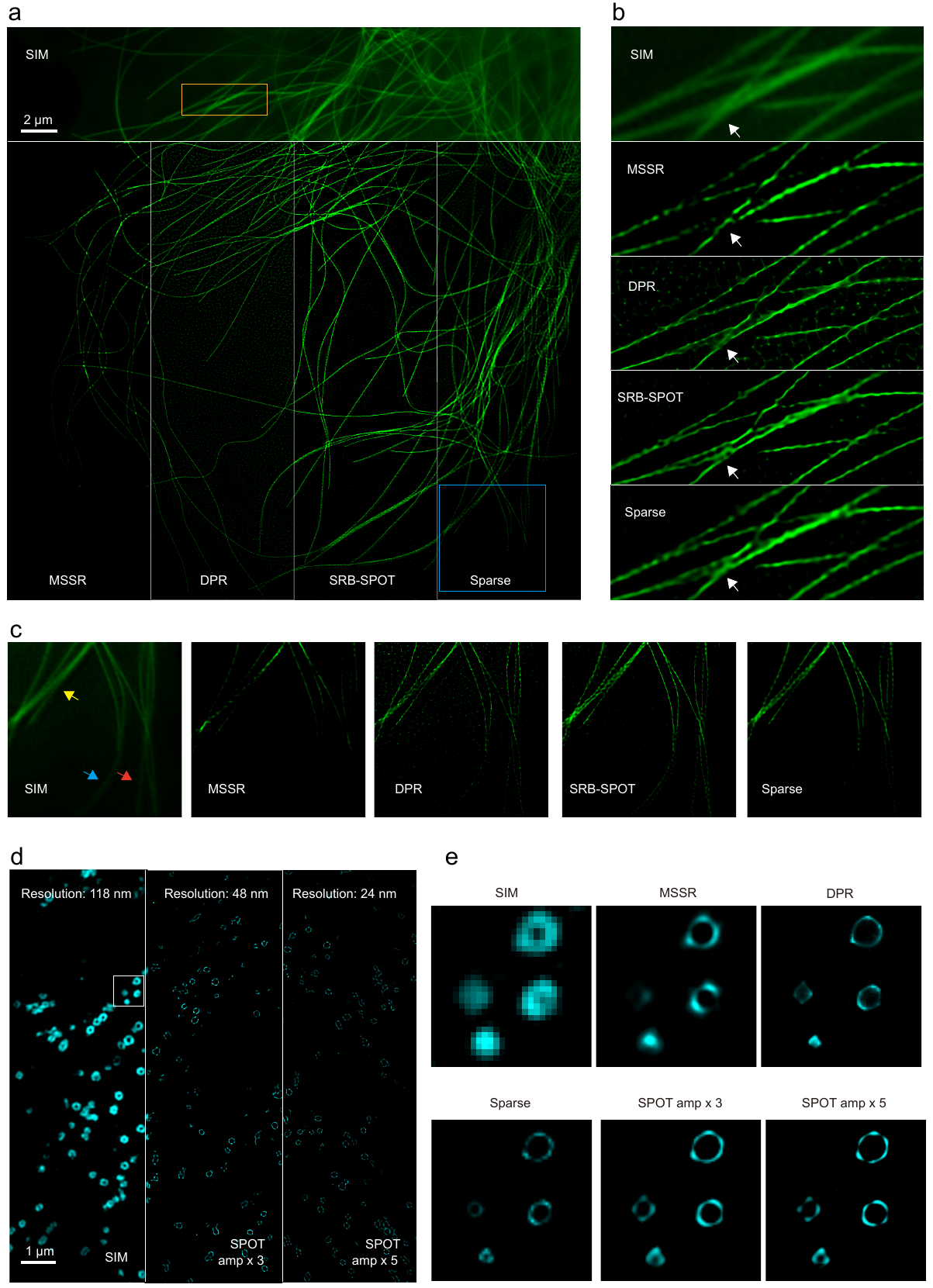}
\caption{
\textbf{Comparison of SPOT and other SR algorithms using
public datasets\cite{qiao2023rationalized,qiao2021evaluation}.}
\textbf{a}, The microtubule raw image was acquired from SIM microscopy and processed using MSSR, DPR, SRB-SPOT, and Sparse Deconvolution\cite{torres2022extending,zhao2023resolution,zhao2022sparse}. 
\textbf{b}, Enlarged views of the orange-boxed region in \textbf{a}. White arrows
highlight structures more clearly resolved by SRB-SPOT.
\textbf{c}, Enlarged views of the blue-boxed region in \textbf{a}. Colored arrows indicate weak-signal regions that are preserved and enhanced by SRB-SPOT.
\textbf{d}, SPOT-processed SIM image of
clathrin-coated pits (CCPs) under amp of 3 and 5. Resolution was estimated
using decorrelation analysis\cite{descloux2019parameter}.
\textbf{e}, Enlarged view of the white-boxed region in \textbf{d},
comparing outputs from different SR algorithms.}
\label{T5}
\end{figure}

\begin{figure}[!htbp]
\centering
\includegraphics[width=0.9\textwidth]{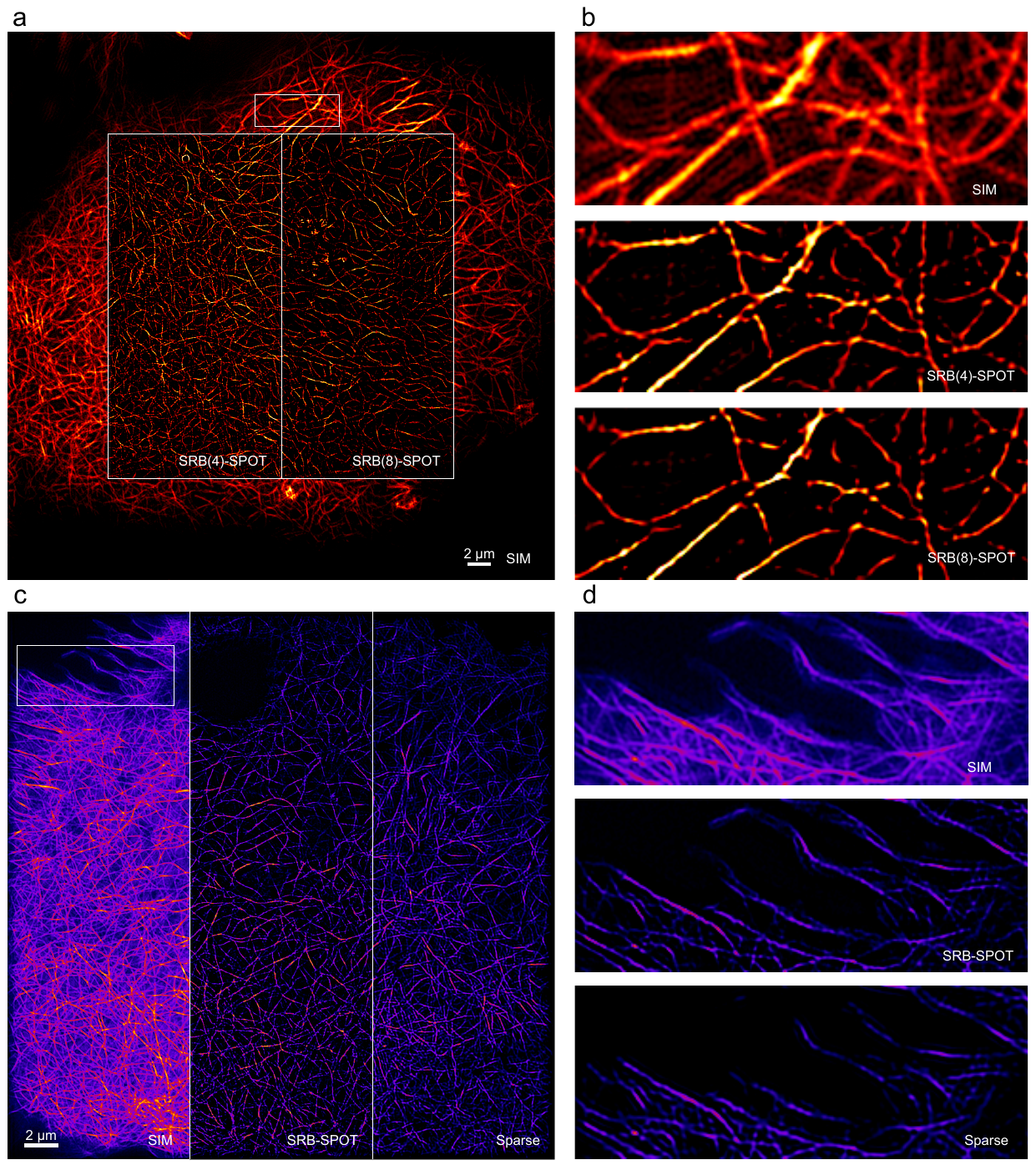}
\caption{
\textbf{Comparison of SPOT and Sparse Deconvolution
algorithms on actin structures using public datasets\cite{qiao2023rationalized,qiao2021evaluation,zhao2022sparse}.}
\textbf{a}, The actin raw image was acquired from SIM microscopy, and processed using
SRB-SPOT with segmentation factors of 4 and 8.
\textbf{b}, Enlarged views of the
white-boxed region in \textbf{a}, showing the effects of different segmentation factors in
SRB-SPOT.
\textbf{c}, The actin raw image was acquired from SIM microscopy, and
processed using SRB-SPOT and Sparse Deconvolution.
\textbf{d}, Enlarged views of the
white-boxed region in \textbf{c}, demonstrating that SRB-SPOT more effectively
suppresses background fluorescence and preserves fine structural details. Scale
bars: 2 $\mu m$.}

\label{T6}
\end{figure}

\begin{figure}[!htbp]
\centering
\includegraphics[width=0.9\textwidth]{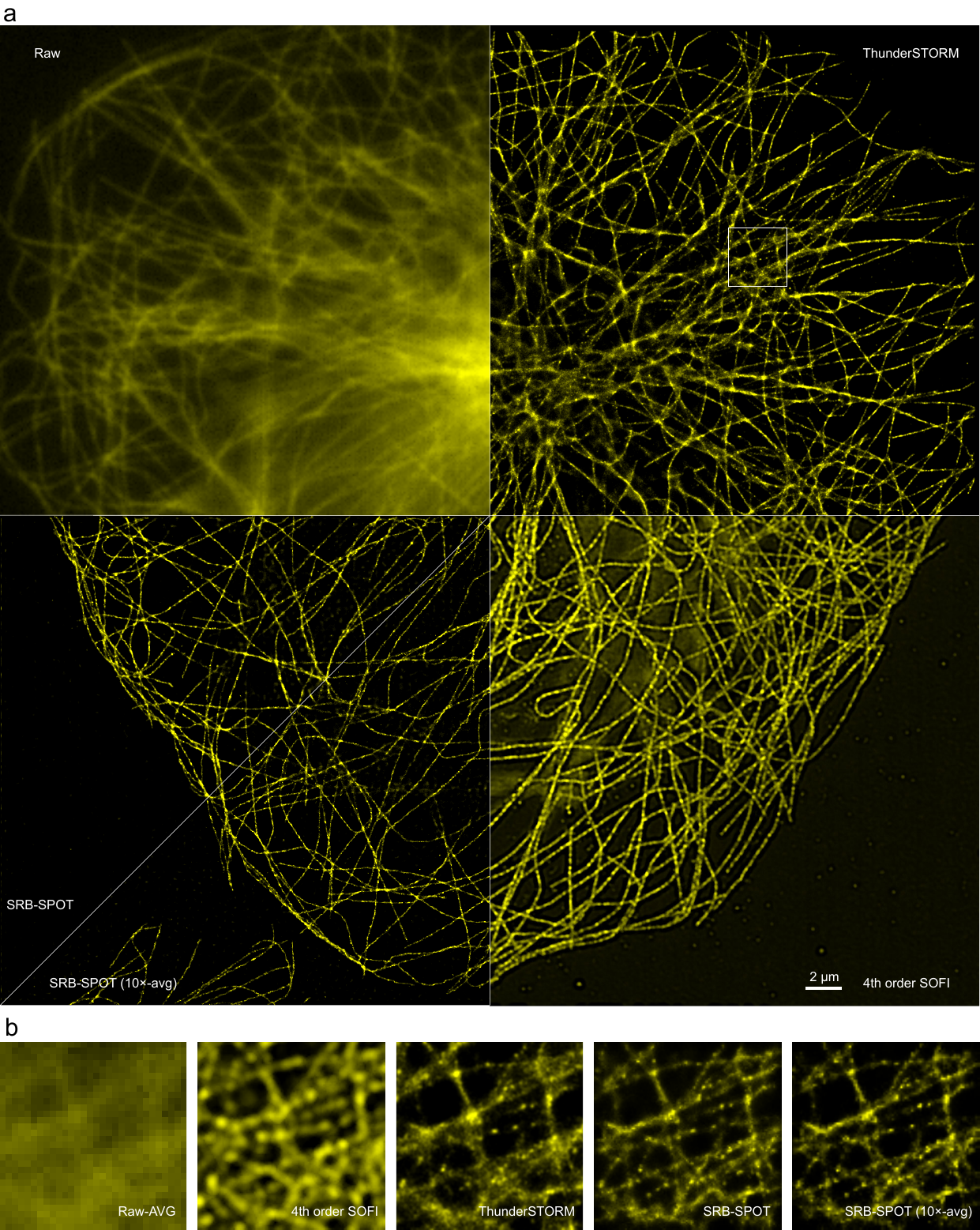}
\caption{
\textbf{Comparison of SPOT with SOFI and ThunderSTORM
using public Tubulin-COS7 dataset\cite{grussmayer2020self}.}
\textbf{a}, The dataset consists of 8000
frames, was processed using ThunderSTORM, SRB-SPOT, SRB-SPOT
(10×-avg), and 4th-order SOFI\cite{dertinger2009fast}. SRB-SPOT (10×-avg) refers to applying
SRB-SPOT to 800 images obtained by averaging every 10 frames. This averaging
simulates a practical reduction in frame rate. The raw
reference image, SRB-SPOT, SRB-SPOT (10×-avg) were generated from the average of their respective image stacks.
\textbf{b}, Enlarged
views of the white-boxed region in \textbf{a}, demonstrating that SRB-SPOT achieves
better structure reconstruction under high-density conditions. Even with
only one-tenth of the original frame number, the SRB-SPOT (10×-avg) result
preserves fine structural details, compared to other methods. Scale bars: 2 $\mu m$.}
\label{T7}
\end{figure}

\end{document}